\documentstyle[12pt]{article}

\begin{document}
%
\newcommand{\beq}{\begin{equation}}
\newcommand{\eeq}{\end{equation}}

\newcommand{\NBI}{\em The Niels Bohr Institute
                  \\ \em University of Copenhagen \\
                     \em  Blegdamsvej 17, DK-2100 \\
                     \em Copenhagen, Denmark}
%
%
\newcommand{\NPB}[3]{{\em Nucl. Phys.} {\bf B#1} (19#2) {#3}. }
\newcommand{\PRD}[3]{{\em Phys. Rev.} {\bf D#1}  (19#2){#3}.  }
\newcommand{\PLB}[3]{{\em Phys. Lett.} {\bf B#1} (19#2) {#3}.   }
\newcommand{\PRL}[3]{{\em Phys. Rev. Lett.} {\bf #1}  (19#2)  {#3}. }
\newcommand{\PR}[3]{{\em Phys. Rev.} {\bf #1}  (19#2) {#3}. }
\newcommand{\CMP}[3]{{\em Commun. Math. Phys.} {\bf #1}  (19#2)  {#3}. }
%
%
%
\pagestyle{empty}

{\hfill \parbox{7cm}{\begin{center} NBI-HE-92-74 \\
                                   October 1992
                    \end{center}}
\vspace{1.5cm}

\begin{center}
\large{\bf Differential Renormalization of a Yukawa Model with
$\gamma_5$}
\end{center}
\vskip .6truein
\centerline {Cristina Manuel
            \footnote[1]{Permanent address: Dept.\ Estructura i
Constituents de la Mat\`{e}ria, Univ.\ de Barcelona, Diagonal 647, 08028
Barcelona (Spain). Bitnet PALAS@EBUBECM1 }} }
\begin{center}
\NBI
\end{center}
\vspace{1.5cm}

\centerline{\bf Abstract}
\medskip
We present a two-loop  computation of the beta functions and the anomalous
 dimensions
of a $\gamma_5$-Yukawa model  using differential renormalization.
The calculation is carried out in coordinate space without modifying the
space-time dimension and no ad-hoc prescription  for
$\gamma_5$ is needed.
 It is shown that this procedure is specially suited
for theories involving $\gamma_5$, and it should be considered in analyzing
chiral gauge theories.

\newpage
\pagestyle{plain}

\section*{I. Introduction}

The differential renormalization  (DR) method \cite{Freedman} has proven
to be a simple and successful way to treat ultraviolet divergences in
a renormalizable quantum field theory
\cite{Haagensen},\cite{Latorre},\cite{RMT},\cite{ji}.
The method is developed in coordinate space and does not change
the dimensionality of space-time. It consists of two steps. In
the first one bare amplitudes, which are too singular at
short distances  to admit a Fourier transform, are
replaced  with derivatives of less singular functions.
In the second one, derivatives are integrated by parts. In doing the
latter, ill-defined surface integrals arise: they correspond to
counterterms that guarantee finite (renormalized) Green functions and
unitarity \cite{Vilasis}. Thus, by discarding surface integrals, one
simultaneously regularizes and
renormalizes  divergent amplitudes.

One of the motivations to develop DR  was to find a consistent
way of treating theories with
 dimension-specific objects \cite{Freedman}.
Theories involving
$\gamma_5$ are of this type and are going to be the
subject of this letter.

Dimensional regularization is the most practical and better
understood regularization method used in quantum field theory but it runs into
difficulties with theories involving $\gamma_5$. There is a
large amount of literature and controversy (see \cite{Bueno} for a review)
 over the correct prescription for $\gamma_5$
in dimensional regularization. There are
basically three different ways to deal with  $\gamma_5$ when
dimensional regularization of the Feynman integrals is
employed, each one of them giving rise to a different
regularization scheme.
The first one is called naive dimensional regularization \cite {Chan}
and  uses an anticommuting
$\gamma_5$ and the  4-dimensional Dirac algebra. The second one
goes under the name of dimensional
reduction \cite{Siegel}, where both the Dirac algebra and the vector
fields are regarded 4-dimensional and split into   D-dimensional bits
plus 4-D scalars. The third one is the
 't Hooft-Veltman proposal \cite{t'Hooft} and
uses a  non-anticommuting $\gamma_5$ in D dimensions.
It has been shown that the only systematic, uniquely fixed and
consistent regularization procedure is the original prescription of \,
`t Hooft and Veltman \cite{Breit},\cite {Bonneau},\cite{Bueno}.
The other two schemes present manifestly algebraic inconsistencies;
they may work for some low-loop calculations, specially when
there are no  closed odd parity fermion loops in the theory, but the
validity of the computations to all loop orders is not guaranteed. However,
in the `t Hooft-Veltman prescription the computational simplicity of
dimensional regularization is lost
and (D-4)-dimensional or evanescent counterterms
have to be taken into account to obtain the correct results. This scheme
is rarely used due to these computational difficulties.

It is our aim to show that DR provides a convenient way to handle
theories with $\gamma_5$. Since in DR the dimensionality of space-time
is kept unchanged, the 4-dimensional Dirac algebra and the standard $\gamma_5$
definition are maintained. As a result, the
algebraic consistency is ensured. The simplicity
of the computations is also another salient feature of DR.

As an example,  we will study a massless Yukawa model with lagrangian
\beq
{\cal L}= \frac{1}{2} (\partial \varphi)^2 - \frac{\lambda}{4!}
\varphi^4+ i \overline{\psi} \gamma \cdot \partial \psi + ie \varphi
\overline{\psi} \gamma_5 \psi.
\label{Lagran}
\eeq
This model has been analysed in the literature using dimensional
regularization  with two of the above prescriptions, namely,
the naive dimensional regularization \cite {Vladi}, and the `t
Hooft-Veltman prescription \cite{Sch}. Here we use DR to compute
the beta functions and the
anomalous dimensions up to two loops  thus avoiding the mentioned
difficulties inherent to dimensional
regularization.
 In sections II and III  some details of the calculations are carefully
explained. Section IV shows the results and  section V is devoted to  the
discussion.

\section*{II. One-loop order}

We will start by studying the model (\ref{Lagran}) at one-loop. This
will give us the chance to recall the
basic techniques used in DR.

We will work in Euclidean space, the Euclidean Feynman rules and some
other conventions being given in Figure 1. Figure  2 shows the 1PI
divergent graphs of the model at one and two loops.

The bare one-loop contribution to the fermion self-energy (Fig.\ 2.1) is
given by
\beq
S_1 ^{bare} (x,y) = - \frac {(ie)^2}{(4\pi^2)^2} \gamma_5
\rlap/\partial_x \frac{1} {(x-y)^2} \gamma_5 \frac{1} {(x-y)^2} = -
\frac {g}{2\pi^2} \rlap/\partial_x \frac{1} {(x-y)^4},
 \label{b1fermion}
\eeq
where (let us emphasize) the standard 4-dimensional Dirac algebra has been
used, and $g \equiv \frac{e^2}{16 \pi^2}$. From
now on, we will use the convention that derivative operators act on the
first variable of the fraction, so that
$\partial_i (x-y)^{-2} = - \partial_i (y-x)^{-2}$.
Clearly, the singularity of the bare amplitude (\ref{b1fermion})  lies in the
$1/(x-y)^4$ factor, which diverges too strongly when $x \rightarrow y$
so as to have a  Fourier transform. Let us consider for a moment $x \neq
y$. Then, the equation \cite{Freedman}
 \beq
\frac{1}{(x-y)^4} = - \frac{1}{4} \Box \frac{\ln
(x-y)^2M^2}{(x-y)^2},
\label{3}
\eeq
is an identity. Now comes the key point: DR defines the renormalized
amplitude as the result of using Eq.\ (\ref{3}) (also at $x = y$) with
the proviso that, when computing the amplitude in momentum space,
the derivatives
acting on the r.h.s. above are formally integrated by parts and the
additional prescription that surface integrals are discarded.
The constant M  has dimensions of mass and plays the role of a
 subtraction point. More explicitly, Eq.\ (\ref{3}) gives for
 the renormalized one-loop contribution to the fermion self-energy:
\beq
S_1 (x,y) =  \frac{g}{8\pi^2} \rlap/\partial \Box
\frac{\ln (x-y)^2 M^2 } {(x-y)^2}.
 \label{r1fermion}
\eeq
Now we can compute  the amplitude in momentum space
(setting y=0 due to translational
invariance)
\beq
\hat{S_1} (p) = \int d^4x e^{ipx} \left(
  \frac{g}{8\pi^2} \rlap/\partial \Box
 \frac{\ln x^2 M^2 } {x^2}
 \right).
 \label{fourier}
\eeq
Following DR,  we  integrate by parts discarding the surface term
 and are left with
\beq
\hat{S_1} (p)= -\frac {ig}{2} \rlap/p \ln \frac {p^2}{4M^2
\gamma^{-2}},
\eeq
where $\gamma= 1.781072...$ is the Euler constant.

The bare one-loop boson self-energy (Fig.\ 2.2) is
\beq
\triangle_1 ^{bare} (x,y) =
-\frac {(ie)^2}{(4\pi^2)^2} tr \left\{\gamma_5 \rlap /\partial
\frac{1} {(x-y)^2} \gamma_5 \rlap /\partial \frac{1} {(y-x)^2}
\right\}= \frac{16g}{\pi^2} \frac{1}{(x-y)^6}.
\label{b1boson}
\eeq
The singular factor $1/(x-y)^6$ in (\ref{b1boson}) now produces, upon
Fourier transform, a quadratic divergence.  To cure it,
 we have to extract two
 derivatives more than in (\ref{3}).  To  this end, we use (see
Appendix  A of ref. \cite{Freedman})
\beq
\frac{1}{(x-y)^6} = -\frac {1}{32} \Box \Box \frac{\ln
(x-y)^2M^2}{(x-y)^2},
 \label{6}
\eeq
thus obtaining the renormalized amplitude
\beq
\triangle_1 (x,y) = -\frac{g}{2\pi^2}\Box\Box\frac{\ln
(x-y)^2M^2}{(x-y)^2}.
 \label{r1boson}
\eeq
 We could compute now the renormalized amplitude
in momentum space, as in the previous case.

We next renormalize the one-loop Yukawa vertex (Fig.\ 2.3). The bare
amplitude is
\begin{eqnarray}
V_1^{bare}  (x,y,z) & = & \frac{(ie)^3}{(4\pi^2)^3} \gamma_5
\rlap/\partial \frac{1} {(y-x)^2} \gamma_5
\rlap/\partial \frac{1} {(x-z)^2} \gamma_5 \frac{1}{(y-z)^2}
\nonumber \\
   &  \equiv &   -ie \frac {g}{4\pi^2} \gamma_5  F(x,y,z).
\label{b1vertex}
\end{eqnarray}
To regulate this graph we notice that the
function $F$ has a singular region when $x \sim y \sim z$. Outside this
region we can integrate by parts and obtain
\begin{eqnarray}
F (x,y,z)  &=& \gamma_i \gamma_j \frac {\partial}{\partial x^i} \left
\{\frac {1}{(x-y)^2 (y-z)^2} \frac {\partial}{\partial x^j}
\frac{1}{(x-z)^2} \right\} \nonumber \\
 & - &  \gamma_i\gamma_j \frac {1}{(x-y)^2 (y-z)^2}
\frac{\partial^2}{\partial x^i \partial x^j} \frac{1}{(x-z)^2}.
\label{fb1vertex}
\end{eqnarray}
The first term on the r.h.s. has a well-defined Fourier transform, as
 can be seen
using dimensional analysis. In turn, the second term is singular at
$x \sim y \sim z$; notice that it contains the laplacian
$\gamma^i \gamma^j \partial_i \partial_j$ and that
\beq
\Box \frac{1}{(x-z)^2} = -4\pi^2 \delta ^{(4)} (x-z),
\label{delta}
\eeq
so the second term in (\ref{fb1vertex}) can be renormalized using
(\ref{3}). Finally we can write the DR renormalized graph as
\begin{eqnarray}
V_1 (x,y,z) & = &- \frac {ieg}{4\pi^4} \gamma_5 \left\{\gamma_i \gamma_j
\frac {\partial}{\partial x^i} \left
(\frac {1}{(x-y)^2 (y-z)^2} \frac {\partial}{\partial x^j}
\frac{1}{(x-z)^2} \right) \right. \nonumber \\
 & - &  \left. \pi^2 \delta^{(4)}(x-z) \Box
\frac{\ln(x-y)^2M^2}{(x-y)^2} \right\}.
\label{r1vertex}
\end{eqnarray}

The 4-point boson function has two contributions at one-loop: the bubble
(Fig.\ 2.4) and the box (Fig.\ 2.5).
The bare bubble is
\beq
\Gamma_{1a} ^{bare} (x,y,z,w) = \frac{\lambda^2}{32 \pi^4} \left [
\delta^{(4)} (x-y) \delta^{(4)} (z-w) \frac{1}{(x-z)^4} + permutations
\right],
 \label{bubble}
\eeq
and  is easily DR renormalized using (\ref{3}):
\beq
\Gamma_{1a} (x,y,z,w) = -2h^2 \delta^{(4)} (x-y) \delta^{(4)} (z-w)
\Box \frac{ln (x-z)^2M^2}{(x-z)^2} + permutations,
\label{rbubble}
\eeq
where $h \equiv \frac {\lambda}{16 \pi^2}$.

The bare box

$\Gamma_{1b} ^{bare} (x,y,z,w)  = $
\beq
 = - \frac{(ie)^4}{(4\pi^2)^4} tr \left
\{ \gamma_5 \rlap/\partial \frac {1}{(x-y)^2}
\gamma_5 \rlap/\partial \frac {1}{(y-w)^2}
\gamma_5 \rlap/\partial \frac {1}{(w-z)^2}
\gamma_5 \rlap/\partial \frac {1}{(z-x)^2} \right\}+ perm.
\eeq
is a bit more complicated due to the presence of
indices but the DR program can be straightforwardly carried out using the Dirac
 algebra and identifying the singular factors. In this
case, we integrate by parts twice and use  (\ref{3}) to obtain the
following DR renormalized
amplitude

$\Gamma_{1b} (x,y,z,w) = $
\begin{eqnarray}
& = &  \frac{g^2}{\pi^4} tr \{\gamma_a \gamma_b \gamma_c
\gamma_d \} \frac {\partial}{\partial y^a} \left(\frac{1}{(y-x)^2}
\partial_b \frac{1}{(y-w)^2}
\partial_c \frac{1}{(w-z)^2}
\partial_d \frac{1}{(z-x)^2} \right) \nonumber \\
 & - &  \frac{16g^2}{\pi^2} \delta^{(4)} (y-w) \frac{1}{(y-x)^2}
\frac {\partial}{\partial z^d}\left( \frac{1}{(z-x)^2}
\frac {\partial}{\partial z^d} \frac{1}{(z-w)^2} \right) \nonumber \\
 & + &  16g^2 \delta^{(4)}(y-w) \delta^{(4)} (z-w) \Box \frac{\ln
(x-w)^2M^2}{(x-w)^2} + perm.
\end{eqnarray}

\medskip

This analysis shows that all one-loop UV divergences are cured
by first isolating the singularities and then by using Eqs.\
(\ref{3}) and (\ref{6}), and the integration by parts
prescription. In this process an arbitrary
parameter with dimensions of mass enters in a natural way. In writing
the different renormalized amplitudes, one has no a priori reason to use
the same $M$ for all graphs but we will use a subtraction scheme where all the
$M's$ are equal. This can be done since there are no Ward identities
 in the theory that force us to introduce different mass scales \cite{Rius}.
  This parameter has the
physical meaning of a renormalization scale, as can be explicitly
seen  when the renormalization group equations
are tested on the renormalized amplitudes (see Section IV).

\section*{III. Two-loop order}

For the sake of brevity, we will not discuss here all the graphs of
the
two-loop order but we will present an overview of the computations, which
involve almost the same techniques used in the one-loop order, that is,
isolating the singularities of bare amplitudes and using DR identities.

Two-loop bare amplitudes present two singular regions.
The regularization of these amplitudes is performed
from the subdivergence, treating it in the same way that it was done at
one-loop, to the overall divergence. Overall divergences require
 new DR identities. In our case, we only need
\begin{eqnarray}
 \frac{\ln x^2M^2}{x^4} & = & -\frac{1}{8} \Box \frac{\ln^2 x^2M^2 + 2 \ln
x^2M^2}{x^2}  \nonumber \\
 \frac{\ln x^2M^2}{x^6} & = & -\frac{1}{64}\Box \Box \frac{\ln^2 x^2M^2 +
5 \ln x^2M^2}{x^2}.
\label{2loops}
\end{eqnarray}
In this way two-loop 1PI are renormalized. As an example, we
examine one of the diagrams which gives contribution to the fermion
self-energy (Fig.\ 2.6), with bare amplitude

$ S_{2a}^{bare}(x,y)  = $
\begin{eqnarray}
 & = & \frac{e^4}{(4\pi^2)^5} \gamma_5
\rlap/\partial \frac{1}{(x-y)^2}\gamma_5
\int \frac{d^4u d^4v}{(y-v)^2(x-u)^2} tr \left\{\gamma_5
\rlap /\partial \frac{1} {(v-u)^2} \gamma_5
\rlap /\partial \frac{1} {(u-v)^2} \right\}   \nonumber \\
 & = & - \frac{4g^2}{\pi^6} \rlap/\partial \frac{1}{(x-y)^2}
\int \frac{d^4u d^4v}{(y-v)^2(x-u)^2 (u-v)^6}.
\label{b2afermion}
\end{eqnarray}
The first  singular region of this amplitude is  the corresponding to the
subdivergence, when $u \sim v$, and is cured in the same way that it was
done at the one-loop order
(\ref{r1boson}). The laplacians in front of $\ln (u-v)^2 / (u-v)^2$ are
then integrated by parts, discarding the surface terms. Then we can
perform easily the integrals over the internal points. The second
singular region corresponds to
the overall divergence, when $x \sim y$, and is cured using (\ref{2loops}) to
finally obtain the DR renormalized amplitude
\beq
S_{2a} (x,y) = - \frac{g^2}{8\pi^2} \rlap/\partial \left(\Box \frac
{\ln^2 (x-y)^2M^2 + 3 \ln (x-y)^2 M^2}{(x-y)^2} \right).
\eeq

Amplitudes with well located subdivergences are cured in a
straightforward way,
so we will not discuss them. There are other diagrams which require
a supplementary effort:
those that contain an overlapping divergence.
In those diagrams it is impossible to separate out the
two singular regions. Nevertheless, overlapping divergences do not
represent a serious difficulty in real-space computations, because the
external points of the amplitude can be kept separated until the
regularization of the subdivergences is accomplished.

Let us take as an example of overlapping divergence the
 amplitude corresponding to Fig.\ 2.10,

$\triangle_{2b}^{bare} (x,y)  =  - \frac{e^4}{(4\pi^2)^5}
tr \{\gamma_a \gamma_b \gamma_c \gamma_d \} \times$
\beq
\int du^4 dv^4 \left \{ \partial_a \frac
{1}{(x-u)^2} \partial_b \frac {1}{(u-y)^2}
\partial_c \frac {1}{(y-v)^2}
\partial_d \frac {1}{(v-x)^2}\frac {1}{(v-u)^2}
\right \}.
\eeq
In this amplitude the singular regions are $u \sim v \sim x$ and $u \sim
v \sim y$. To renormalize it we first extract total derivatives over
the external points. We are then left with three terms.
 We notice that in two of these terms there is a laplacian acting on a
fraction that allows us to obtain a delta function  and perform one
 of the integrals over an internal point. These two terms present
singular regions when $u \sim x$ in one case, and
when $u \sim y$ in the other, and they are renormalized using (\ref{3})
first, and (\ref{6}) and (\ref{2loops}) afterwards.
Also the third term can  easily be  renormalized.
After some algebra, the DR renormalized amplitude is expressed as
\beq
\triangle_{2b}(x,y) = \frac{g^2}{2\pi^2} \Box \Box \frac{\ln^2
(x-y)^2M ^2 + \ln (x-y)^2M ^2}{(x-y)^2}.
\label{r2bboson}
\eeq

The very same procedure is used to renormalize
overlapping
divergences of the Yukawa vertex  and of the 4-point
boson function : extract total derivatives and use
the properties of Dirac matrices during the computation. In these
diagrams, and due to the presence of indices, there are some
integrals that are difficult to solve analytically, but
fortunately they are finite, and they do not need to be renormalized. For
our purposes, to find beta functions and anomalous dimensions, we do not
need them in a explicit form.

As an example of overlapping divergence of a vertex,
we next examine the most difficult two-loop Yukawa vertex (Fig.\ 2.16), with
bare amplitude
\begin{eqnarray}
V^{bare}_{2e} (x,y,z)  = - \frac{ie^5}{(4\pi^2)^6} \gamma_5 \gamma_j
\gamma_i \gamma_l \gamma_k \times ~~~~~~~~~~~~~~~~~~~~~~~~~~~~~~~~~~~\\
\int d^4u d^4v  \left \{ \partial_i \frac{1}{(x-u)^2}
\partial_j \frac{1}{(u-y)^2}
\partial_k \frac{1}{(z-v)^2}
\partial_l \frac{1}{(x-v)^2}
\frac{1}{(u-z)^2(y-v)^2} \right \}.
 ~~~
\nonumber
\end{eqnarray}
To proceed,  we first integrate by parts over $\partial/\partial x^l$.
 Power counting
shows that only the integrated part is too singular so as to have
Fourier transform, so we foan our attention only on that part. The
laplacian acting on $1/(x-u)^2$ gives a delta function,  and  we can then
perform  the integral over $u$. The final integral to be analysed is
\begin{eqnarray}
4\pi^2 \gamma_j \gamma_k \frac{1}{(x-z)^2} \partial_j \frac{1}{(y-x)^2}
\int d^4v  \frac{1}{(x-v)^2(y-v)^2}\partial_k \frac{1}{(z-v)^2} & &
\nonumber \\
\equiv 4\pi^2 \gamma_j \gamma_k \frac{1}{(x-z)^2} \partial_j
 \frac{1}{(y-x)^2} \frac{\partial}{\partial z^k} K(z-x,y-x).  & &
\end{eqnarray}
Now $\partial/ \partial y^j$ is integrated by parts. Once again, only
the integrated part needs renormalization. This last one is separated
into traceless and trace parts. The traceless combination of
derivatives is finite, and the trace part
is straightforwardly DR renormalized.
After some algebra, the final result can be expressed as
\newpage
\begin{eqnarray}
V_{2e} (x,y,z) &=& \frac{ieg^2}{16\pi^8} \gamma_5
\Bigg  \{   -\gamma_j
\gamma_i \gamma_l \gamma_k \times
\nonumber \\
\int d^4u d^4v && \hspace*{-10 mm}
\frac {\partial}{\partial x^l} \left ( \partial_i
\frac{1}{(x-u)^2} \partial_j \frac{1}{(u-y)^2}
\partial_k \frac{1}{(z-v)^2}
\frac{1}{(u-z)^2(y-v)^2(x-v)^2} \right )
\nonumber \\
&+& 4\pi^2 \gamma_j \gamma_k \frac {\partial}{\partial y^j} \left ( \frac
{1}{(x-z)^2(y-x)^2} \frac {\partial}{\partial z^k} K(z-x,y-x) \right)
\nonumber \\
&-& 4\pi^2 \gamma_j \gamma_k \frac{1}{(x-z)^2(y-x)^2}
\left (\frac {\partial^2}{\partial y^j \partial z^k} - \frac
{\delta_{jk}}{4} \frac {\partial^2}{\partial y \cdot \partial z} \right)
K(z-x,y-x)
\nonumber \\
&+& 8\pi^4 \left [ \frac{1}{16}
\Box \frac{\ln (x-y)^2M^2}{(x-y)^2}
\Box \frac{\ln (x-z)^2M^2}{(x-z)^2} \right.
\nonumber \\
&+&\frac{1}{4} \frac {\partial}{\partial x^{\mu}} \left(
\Box \frac{\ln (x-y)^2M^2}{(x-y)^2}
\stackrel{\leftrightarrow}{\frac {\partial}{\partial x^{\mu}}}
\frac{1}{(x-y)^2(x-z)^2} \right) \\
&+&\left. \left.  \frac {\pi^2}{8} \delta^{(4)}(x-z)
\Box \frac{\ln^2 (x-y)^2M^2+ 2\ln (x-y)^2M^2}{(x-y)^2} + (y
\leftrightarrow z)  \right ] \right \}. \nonumber
\end{eqnarray}
Although the final expression seems to be somewhat complicated,
specially for taking its Fourier transform,
the computation of $M\partial / \partial M$
 on it, which is what we need to find the beta functions and the
anomalous dimensions,  presents no difficulty.

The remaining overlapping divergences of the Yukawa vertex and of the
4-point boson function can be treated in a similar way, and we will
not explicitly  discuss them here.

\section*{IV. Renormalization group constants}

After the computation of the renormalized amplitudes, it is easy to find
the values of $M \partial/ \partial M$ acting on all the 1PI graphs. We present
them in Fig.\ 3 in a pictorial form.
With those values, it is possible to check that the renormalized
amplitudes satisfy renormalization group equations
\beq
\left ( M \frac {\partial}{\partial M} + \beta_g \frac
{\partial}{\partial g} + \beta_h \frac {\partial}{\partial h}
- n_{\varphi} \gamma_{\varphi} - n_{\psi} \gamma_{\psi} \right)
\Gamma^{(n_{\varphi},n_{\psi})} = 0.
\label{RGE}
\eeq
When we substitute the values of $M \partial / \partial M$ in
(\ref{RGE}), we find the following values of beta functions and
anomalous dimensions
\begin{eqnarray}
\beta_g & = & 10g^2 + \frac{1}{6} h^2g - 4 hg^2 - \frac {57}{2}g^3
\nonumber \\
\beta_h & = & 3 h^2 + 8 hg - 48 g^2 - \frac {17}{3} h^3 - 12
h^2g+28hg^2+384g^3 \nonumber \\
\gamma_{\varphi} & = & 2g + \frac {1}{12} h^2 - 5g^2 ; \qquad
\gamma_{\psi}  =  \frac {1}{2}g - \frac {13}{8}g^2.
\end{eqnarray}
These results coincide
 with the ones given in the literature, where the same computations were
carried out using naive dimensional regularization \cite{Vladi},
and the 't Hooft-Veltman scheme
 \cite{Sch}\footnote{ There is a 2 factor between the beta
functions obtained in \cite{Sch} and our results, due to the use of a
different convention.}.

This model does not present closed odd parity fermion loops, so one
does not expect the naive dimensional regularization to fail here.
Nevertheless, one would prefer to use a procedure which does not break
the algebraic consistency of the computations and such that one could
always rely on its validity. These problems do not arise in \cite{Sch}, but
there the cumbersomeness of the computations, where evanescent counterterms
had to be taken into account, shows that the 't Hooft-
Veltman scheme is almost impractical.

We must remark, once again, that computations using DR are simple
and algebraic consistent.

\section*{V. Discussion}

We have applied DR to compute the beta functions and anomalous dimensions
up to two loops of the Yukawa model defined in Eq.\ (\ref{Lagran}).

DR presents some salient features.
DR is  just a prescription to extend  distributions which are ill-defined
in some points to distributions well-defined in all space-time.
Bare and DR renormalized amplitudes coincide for non-singular points, so
this procedure
represents a minimal change in the theory.

DR is a real-space method, and computations of multiloop integrals can
be done with ease. Let us recall here that the "method of uniqueness"
\cite{Chet},\cite{Tkac},\cite{Kaz} which also works with coordinate
 amplitudes but
uses dimensional regularization, has already demonstrated the
powerful calculational possibilities of space-time regularization
methods. DR also shares that computational simplicity.
The method is specially suited to obtain renormalization group functions
and renormalized amplitudes in configuration space at high loop order.
The computation of these amplitudes in momentum space requires to
Fourier transform and perform some difficult but well-defined integrals.

In theories in which $\gamma_5$ is present, DR preserves the algebraic
consistency of the computations, since it does not alter the
dimensionality of space-time, and the standard 4-dimensional Dirac
 algebra can be
used.

It is worthwhile to mention here that DR has been used in gauge theories
\cite{Latorre}, \cite{RMT}, and that it reproduces correctly the chiral
anomaly \cite{Latorre}. In gauge theories DR enforces the introduction
of several mass scales, and in order to preserve the Ward identities
certain correlations between them are required. The anomaly results when
the Ward identities overconstraint the value of the mass parameters.

There is still no  proof of the validity of DR to all orders of
perturbation theory. However,
it has very recently been shown \cite{Rius} that there is a
 relationship between
dimensional and differential regularization and renormalization
 in low loop graphs, and one expects that one could find a systematic
application of these relations at higher order graphs and then yield
a consistent proof of DR.  One difference between
the two procedures is that DR
does not modify the space-time dimension, and then it
allows for  a natural and consistent treatment of dimension-specific
 theories.

DR  seems to be a suited and natural method to treat theories with
$\gamma_5$ and, with no doubt, deserves further consideration to
evaluate chiral gauge theories.

\vbox{}
\vfill
{\large \bf Acknowledgements}

I am indebted to J.I. Latorre for introducing me to DR and
guiding me in several stages of this work. I would like to thank the
critical reading of the manuscript of F. Ruiz-Ruiz and
J. Ambj\o rn, the valuable help
of P.E. Haagensen and discussions with X. Vilas\'{i}s-Cardona and D.
Espriu. I am specially grateful to R. Tarrach for a very useful
comment.
I would also like to thank the warm hospitality of the Niels Bohr
Institute, where part of this work was done.

 An FPI grant from
 the Ministerio de Educaci\'{o}n y Ciencia is  acknowledged.

\newpage

\section*{Figure Captions}

Figure 1. Euclidean Feynman rules and other conventions used in the
computations.

\noindent
Figure 2. One and two loop 1PI diagrams. In parenthesis, the number of
diagrams of that type.

\noindent
Figure 3. The value of $M\partial / \partial M$ acting in all
the considered 1PI.



\end{document}